\documentclass[12pt]{amsart}
\usepackage{amssymb,amsmath,amsfonts}
\usepackage{amsaddr}

\usepackage{graphicx,psfrag,epsfig,color}
\usepackage{setspace}
\usepackage{bm}

\usepackage[left=1.0in, right=1.0in, top=1.0in, bottom=1.0in,
  includehead,includefoot]{geometry}

\numberwithin{equation}{section}

\begin{document}

\title[GTPase / ECM feedbacks modulate cellular behavior]{A mathematical model coupling polarity signaling to cell adhesion explains diverse cell migration patterns}

\author{William Holmes}
\address{Department of Physics and Astronomy, Vanderbilt University}

\author{JinSeok Park and Andre Levchenko}
\address{Department of Biomedical Engineering, Yale University}

\author{Leah Edelstein-Keshet}
\address{Department of Mathematics, University of British Columbia, Vancouver BC, Canada}

\date{\today}


\begin{abstract}

\small
Protrusion and retraction of lamellipodia are common features of eukaryotic cell motility. As a cell migrates through its extracellular matrix (ECM), lamellipod growth increases cell-ECM contact area and enhances engagement of integrin receptors, locally amplifying ECM input to internal signaling cascades. In contrast, contraction of lamellipodia results in reduced integrin engagement that dampens the level of ECM-induced signaling. These changes in cell shape are both influenced by, and feed back onto ECM signaling. Motivated by experimental observations on melanoma cells lines (1205Lu and SBcl2) migrating on fibronectin (FN) coated topographic substrates (anisotropic post-density arrays), we probe this interplay between intracellular and ECM signaling.  Experimentally, cells exhibited one of three lamellipodial dynamics: persistently polarized, random, or oscillatory, with competing lamellipodia oscillating out of phase (Park et al., 2016). Pharmacological treatments, changes in FN density, and substrate topography all affected the fraction of cells exhibiting these behaviours. We use these observations as constraints to test a sequence of hypotheses for how intracellular (GTPase) and ECM signaling jointly regulate lamellipodial dynamics.  The models encoding these hypotheses are predicated on mutually antagonistic Rac-Rho signaling, Rac-mediated protrusion (via activation of Arp2/3 actin nucleation) and Rho-mediated contraction (via ROCK phosphorylation of myosin light chain), which are coupled to ECM signaling that is modulated by protrusion/contraction. By testing each model against experimental observations, we identify how the signaling layers interact to generate the diverse range of cell behaviors, and how various molecular perturbations and changes in ECM signaling modulate the fraction of cells exhibiting each.  We identify several factors that play distinct but critical roles in generating the observed dynamic: (1) competition between lamellipodia for shared pools of Rac and Rho, (2) activation of RhoA by ECM signaling, and (3) feedback from lamellipodial growth or contraction to cell-ECM contact area and therefore to the ECM signaling level.


\end{abstract}

\maketitle

\newpage

\section*{Author Summary}
Cells crawling through tissues migrate inside a complex fibrous environment called the extracellular matrix (ECM), which provides signals regulating motility. Here we investigate one such well-known pathway, involving mutually antagonistic signalling molecules (small GTPases Rac and Rho) that control the protrusion and contraction of the cell edges (lamellipodia). Invasive melanoma cells were observed migrating on surfaces with topography (array of posts), coated with adhesive molecules (fibronectin, FN) by Park et al., 2016. Several distinct qualitative behaviors they observed included persistent polarity, oscillation between the cell front and back, and random dynamics. 
To gain insight into the link between intracellular and ECM signaling, we compared experimental observations to a sequence of mathematical models encoding distinct hypotheses. The successful model required several critical factors. (1)  Competition of lamellipodia for limited pools of GTPases. (2) Protrusion / contraction of lamellipodia influence ECM signaling.
(3) ECM-mediated activation of Rho. A model combining these elements explains all three cellular behaviors and correctly predicts the results of experimental perturbations. This study yields new insight into how the dynamic interactions between intracellular signaling and the cell's environment influence cell behavior.

\newpage

\section{Introduction}


Migrating cells display polarization of many membrane and cytosolic components, and spatially inhomogeneous signaling activity. Cellular polarity can be highly dynamic, displaying random, persistent or even oscillatory patterns \cite{camley2013,zhang2014,fraley2012}. In spite of recently proposed phenomenological models attempting to explain how these polarity patterns can emerge in the absence of graded extracellular cues 
\cite{shao2012,Lavi2016}, we still lack the mechanistic understanding of the dynamic molecular mechanisms underlying the polarity establishment and maintenance over the course of cell migration. Thus, given the complexity of the polarity dynamics, we still do not know if diverse spatio-temporal patterns can be accounted for by the same mechanistic framework, quantitatively embedded in a biochemically informed mathematical model. Having such a framework may assist in interventions aimed at enhancement or inhibition of persistence of cell migration in diverse setting, such as wound healing or aggressive cancer spread. 

Aggressive cancers, such as advanced stage melanoma and glioblastoma multiforme, frequently display persistent cell migration away from the primary tumor site. In the context of melanoma, the invasive tumor spread is associated with several mutations, including the loss of functional expression of PTEN, and the corresponding increase in the activity of the PI3K-AKT signaling pathway. It is not clear how such mutations, affecting the state of the signaling and regulatory networks controlling multiple cellular functions, could influence cellular polarity dynamics and the persistence of cell migration. Migrating cells also frequently relocate to micro-environments that are distinct from those of the tissue of origin. One of the key aspects of cellular micro-environment is the organization and composition of the extracellular matrix. Alteration in the density, orientation and nano-topography of the extracellular matrix fibers and their cleaved fragments have been shown to be instrumental in onset of cellular spread and in defining the direction and persistence of cellular migration \cite{vellinga2016,provenzano2006,balcioglu2016}. Recent analysis suggested that these matrix re-arrangements can be well approximated in experiments, using matrix-mimicking nano-fabricated platforms that allow for controlled variation of the model matrix structure and chemical composition \cite{park2016}. In particular, in our experimental analysis with melanoma cell lines  
we found that individual cells can display diverse polarity patterns when migrating in 
areas of the model matrix with 
various degrees of anisotropy  \cite{JPark2016}.
Having this type of controlled micro-environment can allow one to develop mechanistic models \cite{Holmes_Rev_12} of cell polarity control and to test them by checking for consistency between model predictions and experimental results.

In this study, we focused on testing a set of alternative models against the experimental data obtained for melanoma cell lines of different degrees of invasiveness \cite{miller2006,haass2005}. Advanced stages of melanoma are characterized by one of the most invasive behaviors of any cancer, leading to rapid metastatic spread and dismal survival prognosis. In this stage, transformed melanocytes transition from radial to vertical spread patterns, invading the underlying collagen-rich dermis layer and penetrating the blood vessels. Cell polarity and ensuing cell migration patterns can define the effectiveness of the cell invasion, i.e, initial metastatic steps. The experimental dataset used in our analysis represented classification of cell polarization patterns into random, oscillatory and persistent, in the presence of diverse extracellular cues and pharmacological perturbations targeting specific molecular species implicated in polarization control. 

\subsection{Rho GTPases and extracellular matrix signaling}
  
Rho GTPases are central regulators that control cell polarization and migration \cite{Ridley-01,Ridley-03}, embedded in complex signaling networks of interacting components \cite{Devreotes-15}.  Two members of this family of proteins, Rac1 and RhoA, have been identified as key players, forming a central hub that orchestrates the polarity and motility response of cells to their environment \cite{guilluy2011,Byrne-16}. Rac1 (henceforth ``Rac'') works in synergy with PI3K to promote lamellipodial protrusion in a cell \cite{Ridley-03}, whereas RhoA (henceforth ``Rho'') activates Rho Kinase (ROCK), which activates myosin contraction \cite{maekawa1999}. Mutual antagonism between Rac and Rho has been observed in many cell types \cite{Parri-10,Sailem-14,Byrne-16}, and accounts for the ability of cells to undergo overall spreading, contraction, or polarization (with Rac and Rho segregated to front versus rear of a cell). 

The extracellular matrix (ECM) is a jungle of fibrous and adhesive material that provides a scaffold in which cells migrate, mediating adhesion and traction forces. ECM also interacts with cell-surface integrin receptors, to trigger intracellular signaling cascades. Important branches of these pathways are transduced into activating or inhibiting signals to Rho GTPases.  On one hand, ECM imparts signals to regulate cell shape and cell motility. On the other hand, the deformation of a cell affects its contact area with ECM, and hence the signals it receives. The concerted effect of this chemical symphony leads to complex cell behavior that can be difficult to untangle using intuition or verbal arguments alone. This motivates our study, in which mathematical modeling of GTPases and ECM signaling, combined with experimental observations is used to gain a better understanding of cell behavior, in the context of experimental data on melanoma cells.

There remains the question of how to understand the interplay between genes (cell type), environment (ECM) and signaling (Rac, Rho, and effectors). We and others \cite{Symons2009,Parri-10,Sailem-14,Byrne-16,Holmes_PLoS_12,Levchenko-11,Huang-14,Holmes2016} have previously argued that some aspects of cell behavior (e.g., spreading, contraction, and polarization or amoeboid versus mesenchymal phenotype) can be understood from the standpoint of Rac-Rho mutual antagonism, with fine-tuning by other signaling layers \cite{Lawson2014}. Here we extend this idea to couple Rac-Rho to ECM signaling, in deciphering the behavior of melanoma cells in vitro. 
There are several 
overarching questions that this study aims to address.

\begin{enumerate}

\item How does signaling and cell motility interface with external inputs to the cell? How does the change in cell shape (in protrusion/contraction) affect inputs to the signaling network and thus cell behavior?

\item  Are diverse types of cell migration (random persistent, oscillatory) part of the same overall repertoire, or are they distinct and discrete? 

\item  How do constraints such as limited GTPase availability \cite{Maree-06,Jilkine-07,Mori-08, Holmes2016,LEK_RS-13}, lamellipod competition  \cite{Gauthier2011,Raucher2000} and feedbacks (mutual inhibition, positive feedback) determine the cell behavior \cite{tsyganov2012}.

\item  Can we understand the transition to invasive cancer cells as a shift in basic parameters of the same underlying signaling system?

\end{enumerate}

\subsection{Experimental observations constraining the models}\label{exp_results}

In experiments of Park et al. \cite{JPark2016}  
melanoma cells were cultured on micro-fabricated surfaces comprised of post density arrays coated with fibronectin (FN),  represening an artificial extracellular matrix. The anisotropic rows of posts provide 
inhomogeneous topographic cues along which cells orient.  

In \cite{JPark2016}, cell behavior was classified using the well-established fact that
PI3K activity is locally amplified at the lamellipodial protrusions of migrating cells \cite{weiger2009}. PI3K ``hot spots'' were seen to follow three distinct patterns about the cell perimeters: random (RD), oscillatory (OS), and persistent (PS). These classifications were then associated with 
three distinct cell phenotypes: persistently polarized (along the post-density axis), oscillatory with two lamellipodia at opposite cell ends oscillating out of phase (protrusion in one lamellipod coincides with retraction of the other, again oriented along the post-density axis), and random dynamics, whereby cells continually extend and retract protrusions in random directions. The fraction of cells in each category was found to depend on experimental conditions. 

Here, we focus on investigating how experimental manipulations influence the fraction of cells in different phenotypes. For simplicity, we focus on the polarized and oscillatory phenotypes which can be most clearly characterized mathematically. The following 
experimental observations
are used to test and compare our distinct models of cell signaling dynamics.

\begin{figure}[htb] 
  \centerline{ 
\includegraphics[scale=.5]{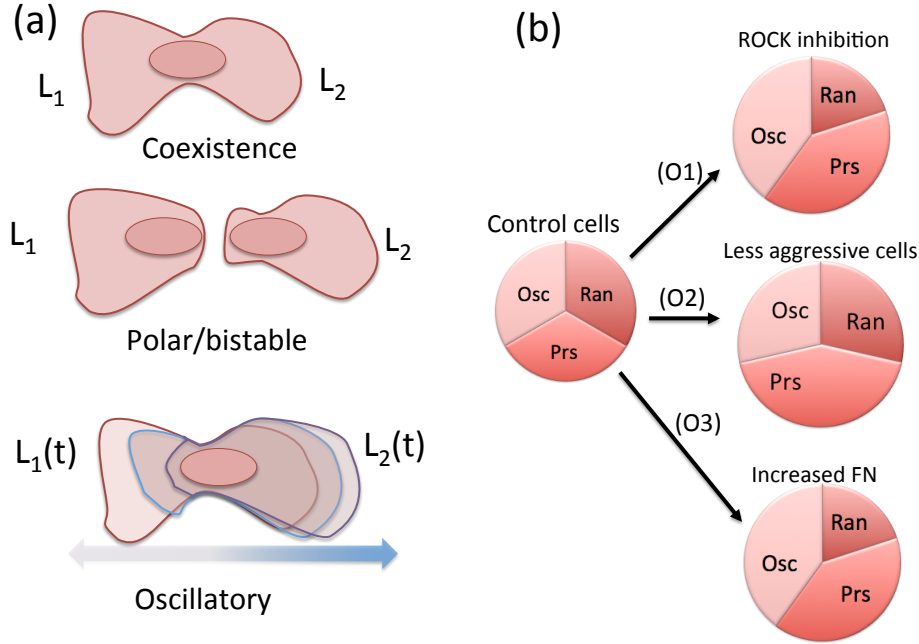}}
\caption{\textbf{Cell phenotypes and experimental observations}: In modeling melanoma cells, we consider signaling and competition of two lamellipodia (indicated by $L_1, L_2$).
\textbf{(a)} Cell states described by our models include (top to bottom) coexistence of two lamellipodia, polarization in which one lamellipod ``wins'' 
the competition (also includes bistability where initial conditions determine the eventual polarity direction), and antiphase oscillations in which one lamellipod expands while the other contracts periodically. The ECM variables $E_1, E_2$ are surrogates for both lamellipod size and ECM signaling activity. \textbf{(b)} (Approximate) experimental observation provide benchmarks agains which proposed models are tested. Relative to control cells, those with either ROCK inhibition or exposed to high fibronectin (FN) level have a lower fraction of randomly directed (apolar) cells, whereas the less invasive cells have a higher fraction of persistently polarized cells.} 
\label{fig:0}
\end{figure}

\begin{itemize}
\item[(O1)] Rho is known to activate Rho Kinase (ROCK), which phosphorylates myosin light chain and leads to actomyosin contraction. Inhibiting ROCK is observed to increase the fraction of polarized and decrease the fraction of oscillatory cells.

\item[(O2)] More invasive melanoma cell lines (1205Lu) are enriched in PI3K and low in the antagonist PTEN. These cells exhibit
a lower fraction of random cells and higher fractions of persistently polarized and oscillatory cells
than the less invasive melanoma cell line SBcl2.

\item[(O3)] 
Increasing fibronectin level on the post density array surfaces increases the fraction of oscillatory and lowers the fraction of persistently polarized cells.

\end{itemize}
For a graphical summary of cell phenotypes and experimental observations, see Figure \ref{fig:0}.

\section{Results}

We use experimental observations (O1-3) as indirect constraints, to distinguish hypotheses for how feedbacks between internal cellular signaling and external ECM inputs modulate cellular behaviors. Toward this end, we construct a collection of simple, predictive models for Rac-Rho-ECM signaling,
and compare model predictions to observations (O1-3) to determine which hypotheses 
are most consistent with experimental results.
The activities of the small GTPases Rac and Rho are used as surrogates for the Rac1/PI3K and RhoA/ROCK pathways that respectively promote actin-based protrusion and acto-myosin based contraction in lamellipodia. 
Based on prior observations \cite{guilluy2011,Huang-14,Byrne-16,Bakal-07,Sailem-14,Cooper-15,Holmes2016}, we assume that Rac and Rho are mutually inhibitory and that at sufficiently high levels, ECM signaling up-regulates activity of RhoA \cite{Danen2002,Park2011}.
While motility regulation is vast with numerous regulators and interactions, we ask what aspects of cell behavior can be explained by this core signaling unit, and take the view that other parts of the signaling cascade serve to fine tune model parameters and inputs.

To compare models to data, we consider three cell states: apolar, persistently polarized, and oscillatory (Fig.~\ref{fig:0}). We interpret these, respectively, in terms of the competition of two lamellipods that can either coexist, exclude one another, or cycle through antiphase oscillations where each grows at the expense of the other \cite{JPark2016}. 
Apolar cells are identified with the ``random'' state described in \cite{JPark2016}, lacking directionality and subject to stochastic fluctuations of polarity (not explicitly modelled).
Each lamellipod is represented as a spatially well-mixed compartment. Hence, our models consist of  ordinary differential equations describing the dynamics and interactions of Rac and Rho within and between these compartments. Moreover, we adopt a caricature of cell-substratum contact area as ECM signaling level, which means that lamellipod size is synonymous with our ECM activity variable. Hence, high Rac activity is assumed to promote ECM signaling (via lamellipod protrusion), whereas high Rho activity has an ECM inhibiting effect (due to lamellipod contraction).` 

\begin{figure}[htb] 
  \centerline{ 
\includegraphics[scale=.8]{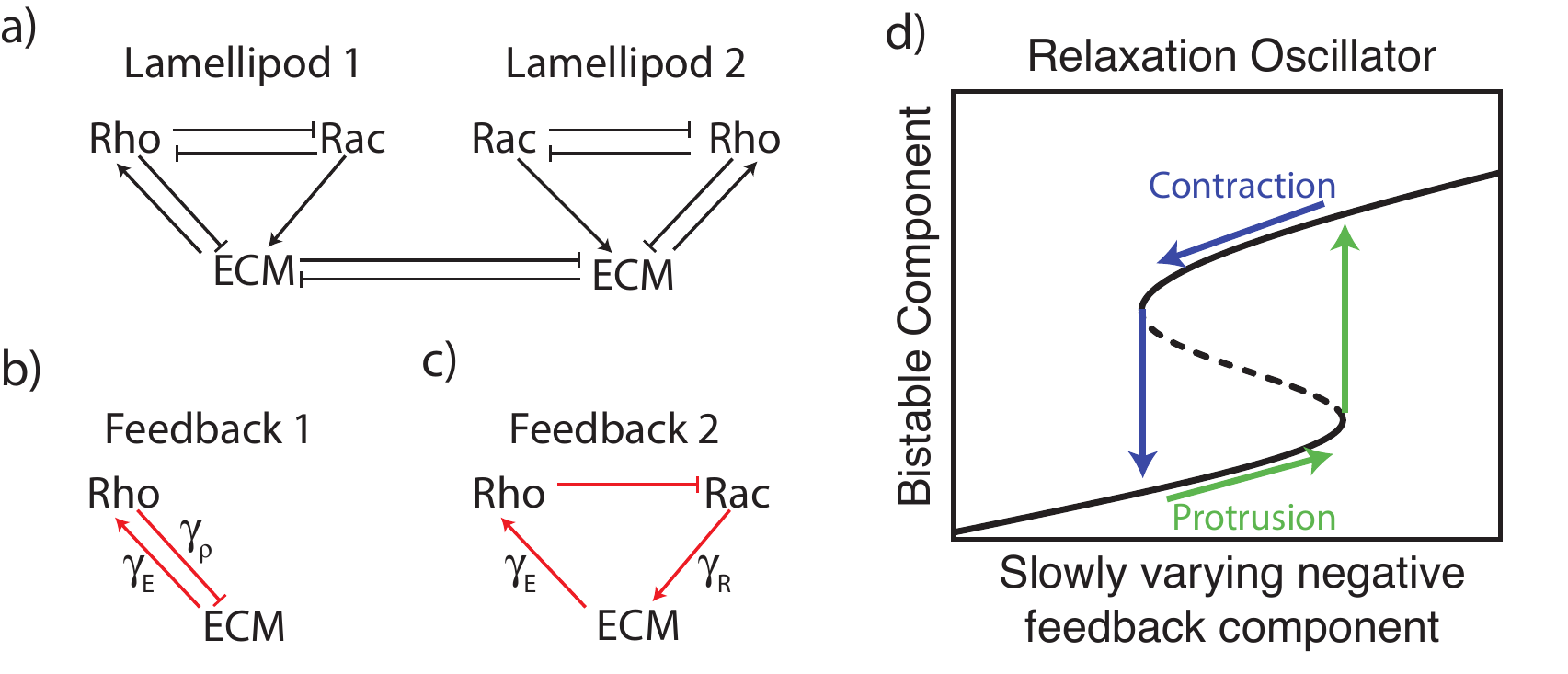}}
\caption{\textbf{Model Schematic:} \textbf{(a)} Schematic depiction of the proposed interactions between the GTPases and extracellular matrix (ECM) signalling.  We assume that
Rac (Rho) promotes protrusion (contraction), resulting in opposite effects on the cell-ECM contact area, and hence on ECM signalling: Rac (Rho) increases (decreases) the level of signalling from the ECM. 
\textbf{(b,c)} We consider two potential negative feedback loops that could lead to oscillations.  Each arrow is labeled with the parameter governing the strength of its influence.  \textbf{(d)} Basic idea of a relaxation oscillator: bistability is present in a subset of the model (black curve); other component(s) acting on a slower timescale provide negative feedback that sets up oscillations (green, blue curves).
In this illustration, bistability results from two fold bifurcations as in the GTPase submodel, but the same idea applies when the bifurcations are both transcritical, as found in the ECM submodel (Fig~\ref{fig:2}d).
   Protrusion causes an increase in ECM signaling that leads to increased feedback, contraction has the opposing effect. 
 }
\label{fig:1}
\end{figure}

 A schematic diagram of the signaling model we discuss is shown in Figure~\ref{fig:1}(a). This overarching model depicts mutually antagonism of Rac and Rho \cite{Symons2009,Parri-10,guilluy2011,Lawson2014,Byrne-16,Holmes2016,Holmes_PLoS_12}, competitive interactions between ECM signaling in the two lamellipodia (assuming growth of one suppresses that of the other), and the influence of ECM signaling on Rho activity.   
 Our models differ by details of the feedbacks and other key assumptions. 
 Figures~\ref{fig:1}(b, c) further illustrate the feedbacks between internal GTPase signaling and ECM signaling layers. 

It is well known that mutual inhibition (or mutual competition) can set up bistability and hysteresis. Furthermore, bistability coupled to slow negative feedback can lead to oscillations \cite{Novak2008}. This idea forms the central theme in our models (Figure~\ref{fig:1}d). 
Moreover, as we will argue, this idea can account for all three observed phenotypes (in appropriate parameter regimes), namely 
a single ``winning'' lamellipod (persistent polarization), apolarity (coexisting lamellipods) and cycling (antiphase oscillations of growth and decay of the two lamellipods). 
The question we address, then, is which subsystem sets up bistability and which leads to oscillations; various interactions between GTPase and ECM signaling levels could, in principle,  account for each. One goal of our modeling is to tease apart the possibilities and find the most likely signaling model that best accounts for experimental observations (Section \ref{exp_results}).

As in most models of intracellular signalling, obtaining biologically accurate values of rate parameters from direct biochemical measurement is unrealistic. Hence, we use the following strategy to parametrize our models. First, we scale the state variables in terms of their inherent ``IC$_{50}$'' levels (levels at which Hill functions are at 50\% of their maximal magnitude). We also scale time by GTPase inactivation times.  This scaling yields a smaller number of ratios of parameters to estimate, i.e. ratios that represent scaled basal and feedback-induced activation rates. For simplicity, we assume that parameters of the Rac and Rho equations are relatively similar.

\subsection{Modeling Overview}

We discuss three model variants, each composed of (A) a subsystem endowed with bistability, and (B) a subsystem responsible for negative feedback. In short, Model 1 assumes ECM competition for (A) and feedbacks mediated by GTPases for (B). In contrast, in Model 2  we assume GTPase dynamics for (A) and ECM mediated feedbacks for (B). Model 3  resembles Model 2, but further assumes limited total pool of each GTPase (conservation), which turns out to be a critical feature. See Tables~\ref{Tab:Source} and \ref{Tab:Elements} for these and other details.

We analyze each model variant as follows: first, we determine (bi/mono)stable regimes of subsystem (A) in isolation, using standard bifurcation methods. Next, we parameterize subsystem (B) so that its slow negative feedback generates oscillations when (A) and (B) are coupled in the model as a whole. 
For this to work, (B) has to force (A) to transition from one monostable steady state to the other (across the bistable regime) as shown in the 
relaxation loop of Figure~\ref{fig:1}d. This requirement informs the magnitude of feedback components. Although these considerations do not fully constrain parameter choices, we found it relatively easy to then parameterize the models (particularly Models 1b and 3). This implies model robustness, and suggests that broad regions of parameter space lead to behavior that is consistent with experimental observations.

\subsubsection{Rac-Rho signaling}

Our underlying Rac-Rho model follows closely on the well-mixed version described in \cite{Holmes2016}. For each GTPase in each lamellipod, we assume a basic activation-inactivation differential equation of the form
\begin{subequations}\label{ModelFormatGeneral}
\begin{equation}
\frac{dG_k}{dt} =  A_{G} G_{I} - \delta_G G_k, \qquad G=R, \rho.
\end{equation}
Here $G=R, \rho$ represent concentrations of active Rac and Rho, respectively, $G_I=R_I, \rho_I$ represent inactive GTPase levels, and $k=1,2$ (e.g. $R_k$) indicates the level in the $k$'th lamellipod.  We assume the inactive GTPase pool is shared by competing lamellipods, but later incorporate different hypotheses about the size of this pool and the extent to which it is depleted.  In some model variants,  we assume that GTPases are abundant so that $R_I, \rho_I$ are constant, whereas in other variants, the total GTPase is limited and conserved
\begin{equation}
\label{GTPaseConserv}
R_I=R_T-R_1-R_2 ,\quad
\rho_I=\rho_T-\rho_1-\rho_2 ,
\end{equation}
\end{subequations}
where $R_T$ and $\rho_T$ are total average concentrations of the GTPases (see Table~\ref{Tab:ModelTerms2}).

We generally assume that mutual antagonism between Rac and Rho influences the rates of activation $A_{R,\rho}$, whereas the rate of inactivation,  $\delta_{R,\rho}$, is constant. (This choice is largely arbitrary and implies GEF-based crosstalk \cite{Markevich2004,Jilkine-07}.) Typical terms are decreasing Hill functions, as shown in Table~\ref{Tab:ModelTerms2} and Eqs.~\eqref{gen-nd}.
If $n=1$ (i.e. Michaelis Menten kinetics) the Rac-Rho system will be monostable, whereas for $n>1$ and appropriate parameters, the Rac-Rho system is bistable.   

\subsubsection{Extracellular matrix input}

The extracellular matrix (ECM) provides input to GTPase signaling. However, the contact area between the cell and the ECM modulates that signaling since larger or smaller lamellipodia receive different levels of stimuli from the ECM. We define the variable $E_k$ to represent the level of ECM signaling,
here viewed as synonymous with the area of the k'th lamellipod (for $k=1,2$). Then Rac-driven cell protrusion and Rho-driven cell contraction will affect $E_k$ on some timescale $1/\hat{\epsilon}$. The lamellipods also compete for growth. For example, Tony Y.-C. Tsai and Julie Theriot observe that neutrophil-like HL-60 cells confined to ``quasi-2D'' motion under an agarose pad have a constant total projected area, for which protruding and contracting lamellipodia compete (personal communication). This suggests that a reasonable model would be 
\begin{subequations}
\begin{eqnarray}
\frac{dE_{k}}{dt}&=&\hat{\epsilon} \left[\mbox{Protrusion} - \mbox{Contraction}  -  \mbox{Competition} 
 \right] \nonumber \\
&=&\hat{\epsilon} \left[P(R_k,E_{k}) -   C(\rho_k,E_{k}) - l_c E_{k}E_{j}   \right], 
\end{eqnarray}
\end{subequations}
where $l_c$ describes the strength of direct lamellipodia competition ($l_c=0$ indicates no competition).
We consider a combination of basal and self-enhanced components of the protrusion term $P$, as shown in Table~\ref{Tab:ModelTerms2}. The Rho-dependent contraction term also captures competition of the lamellipods for growth. 
We will refer to $E_k$ interchangeably as ``ECM signaling level'' and ``lamellipod size'' in our models, with the understanding that these two cell features are intimately linked.

\subsubsection{Dimensionless model equations: general case}
We nondimensionalize time by $\delta_\rho$ and GTPase levels by associated Hill function ``IC$_{50}$ parameters'' ${\hat R_0}, {\hat \rho_0}$ (See Appendix) to arrive at a generic model formulation for the signaling dynamics in each lamellipod $k$ ($k=1,2$):
\begin{subequations}\label{gen-nd}
\begin{align} 
\frac{dR_k}{dt} &= A_R  R_I - \delta R_k, \qquad A_R= \frac{b_R}{1+\rho_k^n},\label{Rac1} \\ 
\frac{d \rho_k}{dt} &= A_\rho  \rho_I - \rho_k ,\qquad A_\rho=\frac{b_{\rho}(E_k)}{1+ R_k^n}, \label{Rho1}\\
\frac{dE_{k}}{dt}&=\epsilon \left[(B_E+ A_E E_{k})  -   E_k\left(L_E E_{k} + l_c E_{j}  \right)   \right],  \quad  j \neq k. \label{ECM}
\end{align}
\end{subequations}
Here, $\epsilon, \delta$ are ratios of timescales and $b_R, b_{\rho}$ are dimensionless.  $B_E, A_E$, and $L_E$ represent basal protrusion, auto-amplified protrusion, and contraction strength terms, which can depend on Rac and Rho concentrations. Rather than nondimensionalizing $E_k$, we retain $l_c$, the only parameter with units of $E$. This allows us to easily control the strength of a coupling between the two lamellipods in our simulations. To represent the fact that ECM signaling influences Rho activation \cite{Danen2002,Park2011}, we assume that the basal rate of Rho activation, $b_{\rho}(E_k)$ is ECM-dependent.

Our models differ in the functional forms assumed for $b_{\rho}, B_E, A_E$, and $L_E$. We present and analyze each of these models one by one, to accentuate their differences and motivate model changes that address specific deficiencies. A summary of definitions and terms used in these models appears in Table~\ref{Tab:ModelTerms2} along with a complete summary of model equations in the Appendix.

\subsection{Correspondence with experiments}\label{correspondence}

Parameters associated with rates of activation and/or feedback strengths are summarized in the Appendix.
 The parameters $\gamma_i$ represent the strengths of feedbacks 1 or 2 in Fig.~\ref{fig:1} (b,c).
$\gamma_{R}$ controls the positive feedback (2) of Rac (via lamellipod spreading) on ECM signaling, and $\gamma_{\rho}$ represents the magnitude of negative feedback  (1) from Rho to ECM signaling (due to lamellipod contraction). $\gamma_E$ controls the strength of ECM activation of Rho in both feedbacks (1) and (2). When these feedbacks depend on cell state variables, we typically use Hill functions with magnitude $\gamma_i$, or, occasionally, linear expressions with slopes ${\bar \gamma_i}$. (These choices are distinguished by usage of overbar to avoid confusing distinct units of the $\gamma$'s in such cases.) 

Experimental manipulations in \cite{JPark2016} (described in Section \ref{exp_results}) can be linked to the following parameter variations.

\begin{itemize}

\item  ROCK inhibition treatment (O1) suppresses the link between Rho activity and actomyosin contraction. Hence this inhibition can be identified with reduction of $\gamma_{\rho}$.

\item Invasiveness (O2) is associated with differences in both PTEN and PI3K activity (more invasive cells exhibit more PI3K and less PTEN). 
 Increasing PI3K activity, or increasing net protrusive activity could correspond to increasing $\gamma_{R}$.

\item Increasing fibronectin (FN) density (O3) leads to increased ECM signaling to Rac, and is thus associated with increased $\gamma_E$ (or $\bar{\gamma}_E$).

\item Membrane tension, cytoskeletal availability, mechanical forces, or other resource limitation in the cell can all potentially affect lamellipodial coupling.  These were not perturbed experimentally but can be represented by variation of the coupling parameter $l_c$.

\end{itemize}
In view of this correspondence between model parameters and experimental manipulations, our subsequent analysis and bifurcation plots will highlight the role of feedback parameters $\gamma_{R,\rho,E}$ in the predictions of each model. Rather than exhaustively mapping all parameters, our goal is to use 1 and 2-parameters bifurcation plots with respect to these parameters to check for (dis)agreement between model predictions and experimental observations (O1-O3). This allows us to (in)validate several hypotheses and identify the eventual model (the Hybrid, Model 3) and set of hypotheses that best account for observations.

\subsection{Lamellipod competition (Models {\bf 1})}
\label{LamelCompDescr}

We first investigated the possibility that lamellipod competition is responsible for bistability and that GTPases interactions create negative feedback that drives the oscillations observed in some cells. To explore this idea, we represented the interplay between lamellipodia (e.g.,  competition for growth due to membrane tension or volume constraints), using an elementary Lotka-Volterra (LV) competition model. For simplicity, we assume that $A_E, L_E$ depend linearly on Rac and Rho concentration, and  set $B_E=0$. (This simplifies subsequent analysis without significantly affecting qualitative conclusions.) With these assumptions, the ECM equations~\eqref{ECM} reduce to the well-known LV species-competition model. 

\begin{figure}[htbp]
  \centerline{ 
\includegraphics[scale=.65]{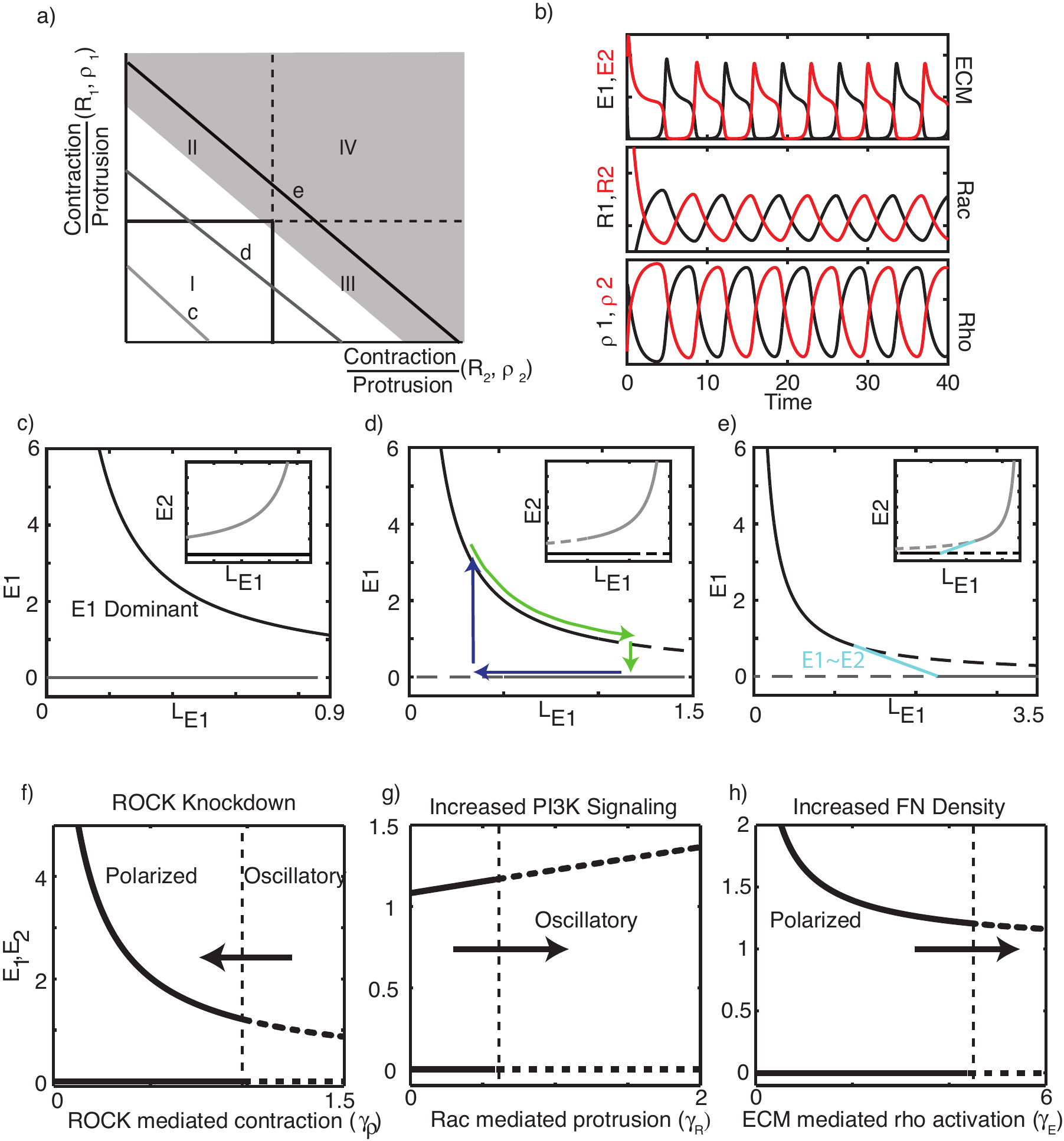}}
\caption{
\textbf{Lamellipod competition (Model 1)}: \textbf{(a)} On its own, the ECM signaling (lamellipod competition) submodel  
(Eqs~\eqref{ECM} with $L_E(\rho_{1,2})$ and $A_E(R_{1,2})$ taken as parameters) has four regimes.
Regimes I-III: a polarized cell with a single lamellipod (bistability in I and competitive exclusion in II-III). Regime IV, lamellipods coexist. Axes represent ratios of $L_E(\rho_k) / A_E (R_k)$  for each lamellipod ($k=1,2$).
 Horizontal and vertical lines are transcritical bifurcations  (at $l_c / A_E(R_k)$. 
\textbf{(b)} Typical simulation of Model \textbf{1b} shows temporal oscillations in ECM signaling
and Rac-Rho activities. $\gamma_{\rho}=1.5$ and all other parameters are as in the Appendix.
\textbf{(c-e)}
One-parameter bifurcation diagrams for $E_1$corresponding to the paths labeled c-e in Panel (a)
(produced by tuning $L_{E1}$ while keeping $L_{E1} +L_{E2}=K$=constant, for $K= 0.9, 1.5, 3.5 $). Other ECM parameters: $A_{E1} = A_{E2} = l_c=1$. Insets show $E_2$ levels. Cycles of protrusion (green) and contraction (blue), shown in panel (d), can occur when feedback from GTPases tunes the relative contraction parameter $L_{E1}$ or $L_{E2}$.
\textbf{(f-h)} Bifurcation diagrams linking predicted behavior to experimental perturbations.
Panel title represents the direction of the arrow on each panel.  Thick curves: stable polar solutions (solid) and oscillations (dashed).
Vertical lines represent Hopf bifurcations. Parameters are as in the Appendix.
 }
\label{fig:2}
\end{figure}

First consider Eqs.~\eqref{ECM} as a function of parameters ($A_E, L_E$), in isolation from GTPase input. As in the classical LV system \cite{Edelstein1988}, competition gives rise to coexistence, bistability, or competitive exclusion, the latter two associated with a polarized cell. These regimes are indicated on the parameter plane of Figure~\ref{fig:2}a with 
the ratios of contractile ($L_E$) and protrusive ($A_E$) strengths in each lamellipod as parameters. (In the full model, these quantities depend on Rac and Rho activities; the ratios $L_E(\rho_k) / A_E (R_k)$ for lamellipod $k=1,2$ lead to aggregate parameters that simplify this figure.) We can interpret the four parameter regimes in Figure~\ref{fig:2}a as follows: I) a bistable regime: depending on initial conditions, either lamellipod ``wins'' the competition. II) Lamellipod 1 always wins. III) Lamellipod 2 always wins. IV) Lamellipods 1 and 2 coexist at finite sizes. Regimes I-III represent strongly polarized cells, whereas IV corresponds to an unpolarized (or weakly polarized) cell.

We next asked whether, and under what conditions, GTPase-mediated feedback could generate relaxation oscillations. Such dynamics could occur provided that slow negative feedback drives the ECM subsystem from an $E_1$-dominated state to an $E_2$-dominated state and back. In Figure~\ref{fig:2}a, this correspond to motion along a path similar to one labeled (d) in Panel (a), with the ECM subsystem circulating between Regimes II and III. This can be accomplished by GTPase feedback, since both
Rho and Rac modulate $L_E$ (contractile strength) and $A_E$ (protrusion strength).
 We show this idea more explicitly in Figure~\ref{fig:2}(c-e) by plotting $E_1$ vs $L_{E1}$ while keeping $L_{E1}+L_{E2}$ constant. (Insets similarly show $E_2$ vs $L_{E1}$.) Each of Panels (c-e) corresponds to a 1-parameter bifurcation plot along the corresponding path labeled (c-e) in Panel (a). 
We find the following possible transitions: In Figure~\ref{fig:2}c, 
we find two distinct polarity states: either $E_1$ or $E_2$ dominate while the other is zero regardless of the value of $L_{E1}$; a transition between such states does not occur. 
 In Figure~\ref{fig:2}d, 
 there is a range of values of $L_{E1}$ with coexisting stable low and high $E1$ values (bistable regime) flanked by regimes where either the lower or higher state loses stability (monostable regimes).
 As indicated by the superimposed loop, a cycle of protrusion (green) and contraction (blue) could then generate a relaxation oscillation as the system traverses its bistable regime. In Figure~\ref{fig:2}e, 
 a third possibility is that the system transits between E1-dominated, coexisting, and E2-dominated states. 
 In brief, for oscillatory behavior, GTPase feedback should drive the ECM-subsystem between
 regimes I, II, and III without entering regime IV.

Informed by this analysis, we next link the bistable ECM submodel to a Rac-Rho system. To ensure that the primary source of bistability is ECM dynamics, a monostable version of the Rac-Rho sub-system is adopted by setting $n=1$ in the GTPase activation terms $A_R, A_\rho$ in Eqs.~\eqref{Rac1}-\eqref{Rho1}. We consider three possible model variants (1a-1c) for the full ECM / GTPase model. 

\subsubsection{Abundant GTPases (Model 1a)}
We first assume that both Rac and Rho are abundant ($R_I, \rho_I$ taken to be constant).  Coupling the GTPase and ECM layers introduces temporal dependence in the parameters $A_E, L_E$ (and thus in the ratios of contraction/protrusion for the lamellipods that form the axes in Figure \ref{fig:2}a). Consequently, a point representing the cell on this figure would drift from one regime to another as the dynamics evolve. In this way, the dynamics of the system as a whole is analyzed from the standpoint of how the GTPase-ECM feedbacks drive the ECM subsystem between its distinct regimes. It follows that oscillating cells are represented by a trajectory that cycles between regimes II and III where the cell would be polarized in opposite directions. 

While in principle, such cycles seem plausible, in practice, we were unable to find them despite reasonable parameter space exploration. (We do not entirely rule out this model, in absence of an exhaustive parameter space exploration and adjustment of all possible kinetic terms.)  Based on our extensive simulations, however, we speculate that oscillations fail for one of two reasons.

\begin{itemize}

\item When the maximum contractile strength is low ($\max{L_E} <  l_c$, which would result from small $\gamma_{\rho}$), it is mathematically impossible for the system to exit Regime I (the border of this regime occurs at $L_E=l_c$). Thus oscillations are not possible since the system cannot enter regions II or III. In this case polarity is possible but not oscillations.

\item When stronger contractile strength is allowed ($\max{L_E} >  l_c$, which would result from larger $\gamma_{\rho}$), the system still does not oscillate. Instead of traversing Regime I, the system crosses into Regime IV. Once in this regime, the apolar solution stabilizes and no oscillation ensues. 

\end{itemize}

As we show next, a small, biologically motivated adjustment discussed in Model variant 1b easily promotes oscillations.

\subsubsection{Limited (conserved) GTPases (Model 1b)}
We next asked how a limited total amount of each GTPase would influence model dynamics. This idea rests on our previous experience with GTPase models in which conservation of the total amount of GTPase played an important role \cite{Maree-06,Jilkine-07,Mori-08,holmes2012regimes,LEK_RS-13,Holmes2016,mata2013model,Holmes-14,Holmes-15}.  To address this question, we augmented Model 1  (Eqs.~\eqref{gen-nd}), with the additional assumption of GTPase conservation \eqref{GTPaseConserv}.  

Conservation introduces a pair of linear algebraic equations of the form $\rho_1+\rho_2+\rho_I=\rho_T$ (and similarly for $R$). Since $L_E(\rho_{1,2})$ depend linearly on $\rho$ in this model variant (Table~\ref{table:equations}), this places a restriction on the sum $L_E(\rho_1)+L_E(\rho_2)$. As a result, part of the parameter space in Figure~\ref{fig:2}a becomes inaccessible (schematically represented in gray) so that the model becomes restricted to the white region. While $L_E(\rho_{1,2})$ could each individually exceed the coupling strength $(l_c)$ (required for oscillations), the system as a whole is unable to access Region IV. In principle, this corrects both issues that led to failure of oscillations in Model 1a, so that slow negative feedback that modulates $L_E(\rho_{1,2})$ should then generate relaxation oscillations. With this adjustment, we indeed found wide parameter regimes corresponding to oscillations (see Figure~\ref{fig:2}b for an example).

We next evaluated this model against experimental observations (O1-O3). First, the model predicts that 
ROCK inhibition suppresses oscillations (Figure~\ref{fig:2}a) while increased fibronectin promotes oscillations (Figure~\ref{fig:2}c), in agreement with (O1) and (O3). Second, increased PI3K (or reduced PTEN which acts as a PI3K antagonist) is linked to Rac-mediated protrusion (Section \ref{correspondence}). This increases the strength of Feedback 2 and promote oscillations (Figure~\ref{fig:2}b). Given the link between PTEN suppression / elevation of PI3K in invasive cells, this is consistent with O2.

While this model can account for all three observations, one significant issue leads us to reject it. The timescales required to generate oscillations in this model are inconsistent with known biological timescales. Relaxation oscillators require that a {\it slow} variable provides the negative feedback that promotes oscillations. Since Model 1 is predicated on negative feedback from GTPases to bistable 
ECM-cell contact area subsystem,
it implies that GTPase dynamics must occur on a slower time scale than that of the cell-ECM subsytem.
This appears to be unreasonable based on the fact that GTPase activation/inactivation operates on a typical timescale of seconds, much faster than the actomyosin-based protrusion and contraction of cells. Indeed, by varying the timescales so that GTPase dynamics are faster than ECM dynamics ($\epsilon < 1$), we find that oscillations no longer occur (results not shown). For this reason, we reject Model 1b as it stands.

\subsubsection{GTPase effectors (Model 1c)}

Before dismissing Model 1 on grounds of timescale, we considered one additional modification. We asked whether the fast timescale of GTPases could be retarded by downstream effectors that participate in relevant
feedback loops. 
To study this possibility, we supplemented Eqs.~\eqref{gen-nd}, \eqref{GTPaseConserv} with dynamics of intermediate Rac effectors $w_{1,2}$  (e.g. WASP, WAVE, PI3K, or other downstream components) and Rho effectors $c_{1,2}$ (e.g. ROCK, etc.) that could correct the timescale problem.
These putative effectors are represented as simple dynamic variables in Eqs.~\ref{full-model-2c} (See Appendix and Figure \ref{fig:5} for a schematic), with the parameter  $\epsilon_2$ governing timescale. In the $\epsilon_2 \gg \epsilon$ limit, this variant reduces (by a quasi steady state approximation) to Model 1b.

The structure of Model 1c and the number and values of steady states are the same as in Model 1b. Only the timescales associated with various model components change. This model could hence account for the same experimental observations as Model 1b. Further, if $\epsilon_2 < \epsilon < 1$, it can do so with GTPase dynamics faster than ECM dynamics. Unfortunately, while GTPase dynamics no longer need be slow, dynamics of ROCK / WASP must be slower than ECM dynamics, which is still biologically implausible.

 To summarize, we have grounds for rejecting a model in which ECM dynamics are central to the generation of bistability and polarity. To achieve oscillations with such a model, other components of the system must operate on a slow timescale.
Biologically, since GTPases and their effectors regulate protrusion and contraction, it is only reasonable that they operate on a faster timescale than lamellipodial dynamics, and cannot therefore be the source of slow negative feedback. This motivates the development of our next attempt in which we reverse the roles of GTPases and ECM as sources of bistability and negative feedback.


\subsection{Bistable GTPases (Model {\bf 2})}

In view of the conclusions  thus far, we now explore the possibility that bistability stems from mutual antagonism between Rac and Rho, rather than lamellipod competition. To do so, we chose Hill coefficients $n=3$ in the rates of GTPase activation, $A_{R}, A_{\rho}$. We then assume that ECM signaling both couples the lamellipods and provides the requisite slow negative feedback.  Here we consider the case that GTPases are abundant, so that the levels of inactive Rac and Rho ($R_I, \rho_I$) are constant.  

\begin{figure}[htb] 
 \centerline{ 
\includegraphics[scale=.9]{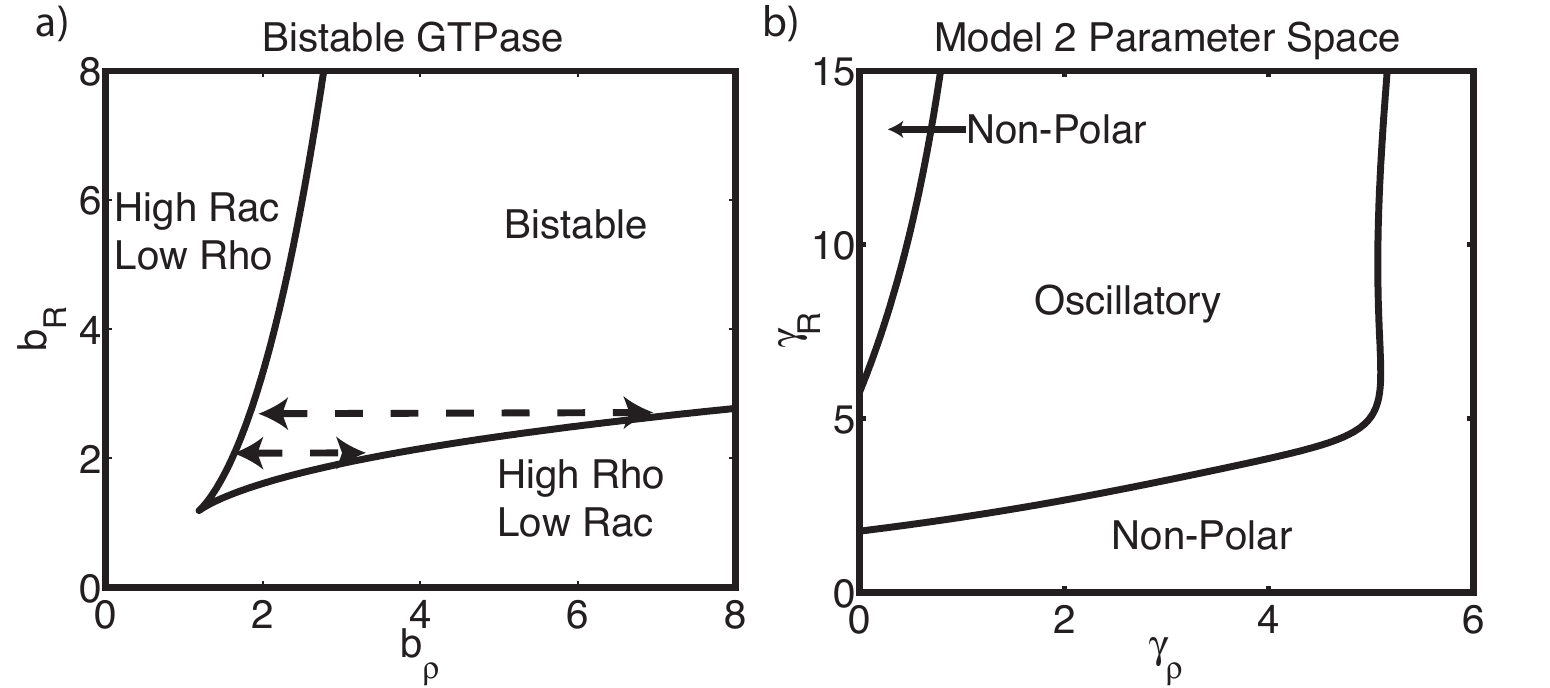}}
\caption{ Bistable GTPases Model {\bf 2}: \textbf{(a)} Bifurcation analysis
of Eqs.~\eqref{Rac1}, \eqref{Rho1} with respect to GTPase activation rate parameters $b_R$ and $b_{\rho}$ (here assumed fixed, independent of $E_k$ to decouple the ECM module). Other parameters: $n=3$, $\delta=1$. Dashed lines indicate the bistable range of $b_{\rho}$ values that must be traversed to induce oscillation. Compare to Figure \ref{fig:4}a.  \textbf{(b)} Bifurcation analysis of Model {\bf 2} with respect to  the strength of the feedbacks $\gamma_{R}$ and $\gamma_{\rho}$.  Two curves of Hopf bifurcations separate oscillatory and static (but non-polar) regimes of behavior.  Parameters values are provided in the Appendix.
}
\label{fig:3}
\end{figure} 

We first characterize the GTPase dynamics with $b_{R,\rho}$ as parameters. Subsequently, we include ECM signaling dynamics and determine how the feedback drives the dynamics in the $(b_R, b_{\rho})$ parameter plane. 

Isolated from the ECM influence, each lamellipod is independent so we only consider the properties of GTPase signaling in one. This mutually antagonistic GTPase submodel is the well-known ``toggle switch'' \cite{Gardner-00} that has a bistable regime, as shown in the  $(b_R, b_{\rho})$ plane of Figure \ref{fig:3}a. ECM signaling affects the Rac / Rho system only  as an input to $b_{\rho}$. A linear dependence of $b_{\rho}$ on $E_k$ failed to produce an oscillatory parameter regime, so we used a nonlinear Hill type dependence with basal and saturating components. Furthermore, for GTPase influence on ECM signaling we use Hill functions for the influence of Rho (in $L_E$) and Rac (in $B_E$) on protrusion and contraction. We set $A_E=0$ in this model for simplicity. (Nonzero $A_E$ can lead to compounded ECM bistability that we here do not consider.) 

Given the structure of the $b_{\rho}-b_R$ parameter plane and the fact that  ECM signaling variables only influence $b_{\rho}$, we can view oscillations as periodic cycles of contraction and protrusion forming a trajectory along one of horizontal dashed lines in Figure \ref{fig:3}a. This idea guides our parametrization of the model. We select a value of $b_R$ that admits a bistable range of $b_{\rho}$ in Figure \ref{fig:3}a. Next we choose maximal and minimal values of the function $b_{\rho}(E_K)$ that extend beyond the borders of the bistable range. This choice means that the system transitions from the high Rac / low Rho state to the low Rac / high Rho state over each of the cycles of its oscillation. With this parametrization, we find oscillatory dynamics, as shown in Figure \ref{fig:3}b. 

We now consider the two-lamellipod system with the above GTPase module in each lamellipod; we challenge the full model with experimental observations. Since each lamellipod has a unique copy of the Rac-Rho module, ECM signaling provides the only coupling between the two lamellipods. First, we observed that inhibition of ROCK  (reduction of $\gamma_{\rho}$ in Figure \ref{fig:3}b) suppress oscillations. However the resulting stationary state is non-polar, in contrast to experimentally observed increase in the fraction of polarized cells (O1). We adjusted the coupling strength ($l_c$) to ensure that this disagreement was not merely due to insufficient coupling between the two lamellipods. While an oscillatory regime persists, the discrepancy with (O1) is not resolved: the system oscillates, but inhibiting ROCK gives rise to a non-polarized stationary state, contrary to experimental observations.

Yet another problematic feature of the model is its undue sensitivity to the strength of Rac activation ($b_R$). This is evident from a comparison of the dashed lines in Figure \ref{fig:3}a. A small change in $b_R$ (vertical shift) dramatically increases the range of bistability (horizontal span), and hence the range of values of $b_{\rho}$ to be traversed in driving oscillations. This degree of sensitivity seems inconsistent with biological behavior. 

It is possible that an alternate formulation of the model (different kinetic terms or different parametrization) might fix the discrepancies noted above, so we avoid ruling out this scenario altogether. In our hands, this model variant failed. However a simple augmentation, described below, addresses all deficiencies, and leads to the final result.


\subsection{Hybrid (Model {\bf 3})}

In our third and final step, we add a small but significant feature to the bistable GTPase model to arrive at a working variant that accounts for all observations. Keeping all equations of Model 2, we merely drop the assumption of unlimited Rac and Rho. We now require that the total amount of each GTPase be conserved in the cell. This new feature has two consequences. First, lamellipods now compete not only for growth, but also for limited pools of Rac and Rho. This, along with rapid diffusion of inactive GTPases across the cell \cite{Postma2004,Jilkine-07,Mori-08} provides an additional global coupling of the two lamellipods. This seemingly minor revision produces novel behavior.

\begin{figure}[htbp] 
  \centerline{ 
\includegraphics[scale=.85]{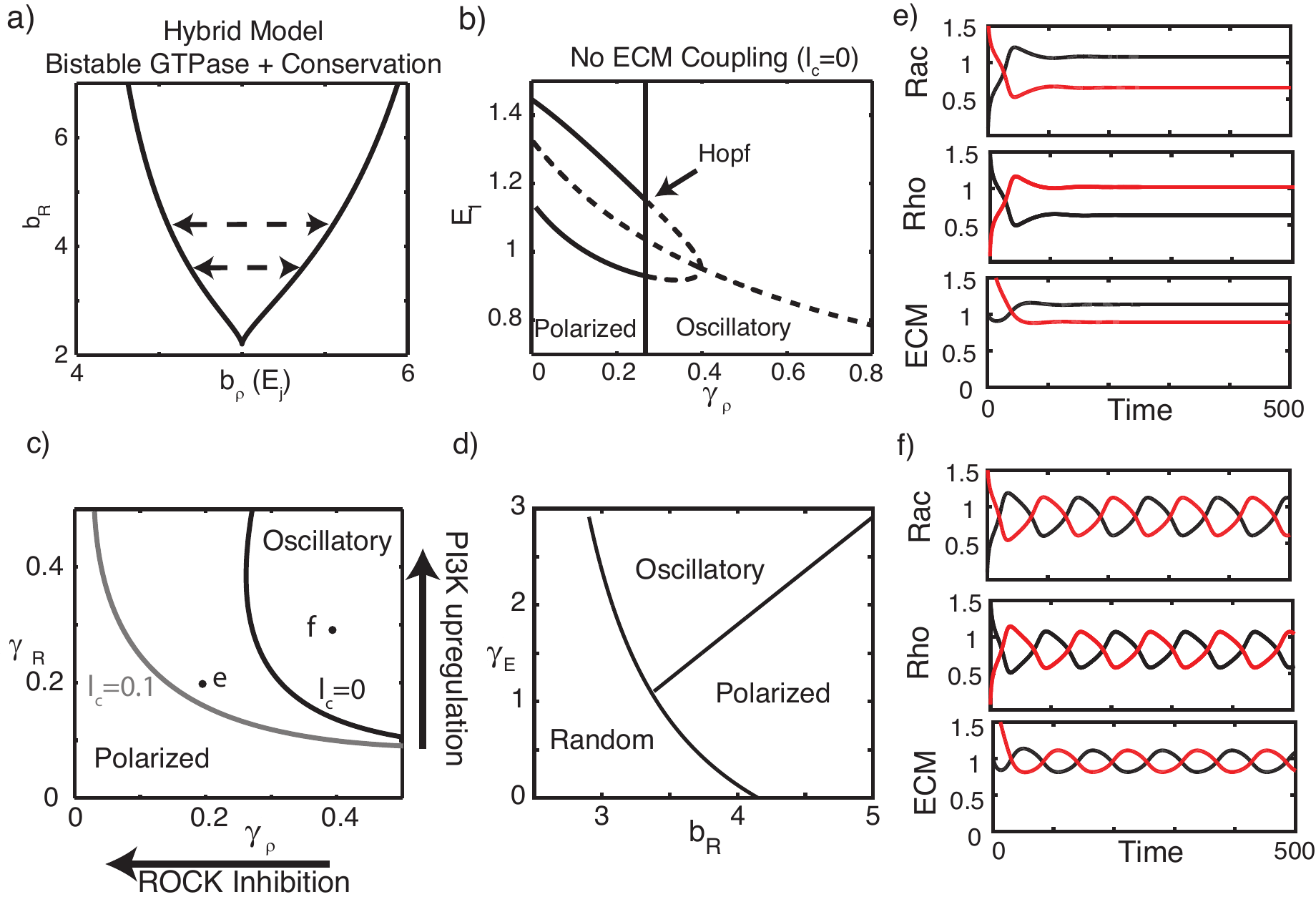}}

\caption{
\textbf{Hybrid Model (Model 3):} Assuming that the pools of Rac and Rho are constant in the cell fixes undesirable features of Model 2.
 \textbf{(a)} As in Fig.~\ref{fig:3}(a) but with GTPase conservation Eqs.~\eqref{GTPaseConserv}. 
Other parameters: $n=3$, $\delta=1, \rho_T=R_T=2$. Dashed lines indicate the range of $b_{\rho}$ values that must be traversed to induce oscillation. 
\textbf{(b)} Bifurcation analysis with respect to $\gamma_{\rho}$. Polarized (indicated by stable steady states, solid curves) and oscillatory (unstable steady states, dashed curves) regimes are present.
\textbf{(c)} Two-parameter bifurcation diagrams depicting oscillatory and polarized parameter regimes separated by a locus of Hopf bifurcations, both with ($l_c=0.1$) and without ($l_c=0$) direct lamellipod coupling.  Points (e,f) indicate parameter values at which simulations were performed (with $l_c=0$) to confirm dynamics.
\textbf{(d)} Bifurcation diagram of the full model, with respect to basal Rac activation rate $b_R$ and ECM-induced Rho activation rate $\gamma_E$, concordant with such diagrams in \cite{JPark2016}. 
\textbf{(e,f)} Sample simulations showing lamellipodial polarity and oscillations (ECM signaling levels in each lamellipod over time). Parameter values correspond to the labeled points in panel (c). For (e), $l_c=0, \gamma_R=0.2, \gamma_{\rho}=0.2$. For (f), $l_c=0, \gamma_{R}=0.3, \gamma_{\rho}=0.4$. All other parameters are given in the Appendix.
}
\label{fig:4}
\end{figure} 

We proceed as before, first analyzing the GTPase signaling system on its own. With conservation, the $b_R- b_{\rho}$ plane has changed from its previous version (Figure \ref{fig:3}a for Model 2) to Figure \ref{fig:4}a. For appropriate values of $b_R$, there is a significant
bistable regime in $b_{\rho}$. Indeed, we find 
three regimes of behavior as the contractile strength in lamellipod $k$, $b_{\rho} (E_k)$,  varies:
a bistable regime where polarity in either direction is possible, a regime where lamellipod $j$ ``wins'' ($E_j > E_k$, left of the bistable regime), and a regime where lamellipod $k$ ``wins'' (right of the bistable regime). Only polarity in a single direction is possible on either side of the bistable regime.

As in Model 2, we view oscillations in the full model as cycles of lamellipodial protrusion and contraction that modify $b_{\rho}(E_k)$ over time, and result in transitions between the three polarity states. To parameterize the model, we repeat the process previously described (choose a value of $b_R$ consistent with bistability, then choose the dependence of $b_{\rho}$ on ECM signaling so as to traverse that entire bistable regime.)

We couple the GTPase system with ECM equations as before. We then check for agreement with observations (O1-O3). As shown in Figure \ref{fig:4} (e,f), the model produces both polarized and oscillatory solutions. To check consistency with experiments, we mapped the dynamics of this model with respect to both ROCK mediated contraction and PI3K mediated protrusion (Figures \ref{fig:4} c). Inhibiting ROCK (Figure \ref{fig:4}b, decreasing $\gamma_\rho$) results in a transition from oscillations to polarized states, consistent with (O1).  
PI3K upregulation promotes oscillations (increasing $\gamma_R$, Figure~\ref{fig:4}c), characteristic of the more invasive cell line 1205Lu (consistent with O2). 
Finally, increased fibronectin density (increased $\gamma_E$, Figure~\ref{fig:4}d) also promotes oscillations, consistent with (O3).  We conclude that this Hybrid Model can account for polarity and oscillations, and that it is consistent with the three primary experimental observations (O1-3). Finally, Model 3 can recapitulate such observations with more reasonable timescales for GTPase and ECM dynamics than were required for Model variant 1b.

It is apparent that Model 3 contains two forms of lamellopodial coupling: direct (mechanical) competition and competition for the limited pools of inactive Rac and Rho. While the former is certain to be an important coupling in some contexts or conditions \cite{diz2016membrane}, we find that it is dispensable in this model (e.g, see $l_c=0$ in Figure \ref{fig:4}c). We comment about the effect of such coupling in the Discussion.


\subsection{Summary of Results}

To summarize, Model 1b was capable of accounting for all observations, but required conservation of GTPase to do so. This model was however rejected due to unreasonable time scales needed to give rise to oscillations. Model 2 could account for oscillations with appropriate timescales, but it appears to be highly sensitive to parameters and, in our hands, inconsistent with experimental observations. Model 3, which combines the central features of Models 1b and 2, has the right mix of timescales, and agrees with experimental observations. In that final Hybrid Model, ECM based coupling ($l_c$) due to mechanical tension or competition for other resources is not essential, but its inclusion makes oscillations more prevalent (Figures \ref{fig:4} b,e).

Additionally, in this Hybrid Model, we identify two possible negative feedback motifs, shown in Figure \ref{fig:1}b. These appear to work cooperatively in promoting oscillations. As we have argued, feedbacks are tuned so that ECM signaling spans a range large enough that 
$b_{\rho}(E_k)$
traverses the entire bistable regime (Figure \ref{fig:4}a). This is a requirement for the relaxation oscillations schematically depicted in Figure \ref{fig:1}c. Within an appropriate set of model parameters, either feedback could, in principle, accomplish this. Hence, if Feedback 1 is sufficiently strong, Feedback 2 is superfluous and vice versa. Alternatively, if neither suffices on its own, the combination of both could be sufficient to give rise to oscillations. Heterogeneity among these parameters could thus be responsible for the fact that in ROCK inhibition experiments (where Feedback 1 is essentially removed), most but not all cells transition to the persistent polarity phenotype.

The Hybrid Model (Model 3) is consistent with observations O1-O3. We can now challenge it with several further experimental tests. In particular, we make two predictions.
\begin{itemize}
\item[(P1)] In the model, reducing the rate of Rac-mediated protrusion ($\gamma_R$) promotes persistent polarization over oscillation (Fig.~\ref{fig:4}e,f). One way to test this experimentally is to inhibit Rac activity. Rac inhibition experiments are reported in \cite{JPark2016}, which validate the prediction: oscillations are suppressed in favor of persistent polarization.
\item[(P2)] Alternatively, the Hybrid model predicts that promoting Rho-mediated contraction (increasing $\gamma_{\rho}$) promotes oscillations over persistent polarization. 
Experimentally, this effect was achieved by applying nocodazole, which breaks up microtubules (MTs) and releases the MT-sequestered RhoGEFs, which subsequently activate Rho and contraction. Experiments of this type reported in \cite{zhang2014,JPark2016}. The latter indeed showed that MT suppression promotes oscillation, consistent with the model prediction.
\end{itemize}


\section{Discussion}

Migrating cells can exhibit a variety of behaviors. These behaviors can be modulated by the cell's internal state, its interactions with the environment, or mutations such as those leading to cancer progression. In most cases, the details of mechanisms underlying a specific behavior, or leading to transitions from one phenotype to another are unknown or poorly understood. Moreover, even in cases where one or more defective proteins or genes are known, the complexity of signaling networks make it difficult to untangle the consequences. Hence, using indirect observations of cell migration phenotypes to elucidate the properties of underlying signaling modules and feedbacks are, as argued here, a useful exercise.

Using a sequence of models and experimental observations (O1-O3) we tested several plausible  hypotheses for melanoma cell migration phenotypes observed in \cite{JPark2016}. By so doing, we found that GTPase dynamics are fundamental to providing 1) bistability associated with polarity and 2) coupling between competing lamellipods to select a single ``front'' and ``rear''. (This coupling is responsible for the antiphase lamellipodial oscillations). Further, slow feedback between GTPase and ECM signaling resulting from contraction and protrusion generate oscillations similar those observed experimentally.

The single successful model, Hybrid Model (Model 3), is essentially a relaxation oscillator. Mutual inhibition between the limited pools of Rac and Rho, sets up a primary competition between lamellipods that produces a bistable system with polarized states pointing in opposite directions. Interactions between GTPase dynamics and ECM signaling provide the negative feedback required to flip this system between the two polarity states, generating oscillations for appropriate parameters. Results of Model 3 are consistent with observations (O1-O3), and lead to predictions (P1-P2), that are also confirmed by experimental observations \cite{JPark2016}. In \cite{JPark2016}, it is further shown that the fraction of cells exhibiting each of these behaviors can be quantitatively linked to heterogeneity in the ranges of parameters representing the cell populations in the model's parameter space.

These results provide interesting insights into the biology of invasive cancer cells (in melanoma in particular), and shed light onto the mechanisms underlying the extracellular matrix-induced polarization and migration of normal cells. First, they illustrate that diverse polarity and migration patterns can be captured within the same modeling framework, laying the foundation for a better understanding of seemingly unrelated and diverse behaviors previously reported. Second,  our results present a mathematical and computational platform that distills the critical aspects and molecular regulators in a complex signaling cascade; this platform could be used to identify promising single molecule and molecular network targets for possible clinical intervention.


\section{Acknowledgments} 
WRH was supported by National Science Foundation grants DMS1562078 and SES1556325. LEK was supported by an NSERC Discovery Grant RGPIN 41870-12. AL and JSP were supported by NIH National Cancer Institute grant U54CA209992.

\bibliography{GTPasebib}

\appendix

\section{Methods}

We simulate models using MatLab's (MathWorks, Natick MA) built-in ODE library (ODE45).  We use dynamical systems bifurcation techniques and the package Matcont \cite{Matcont-03} to map the parameter space  of each model and create bifurcation diagrams.  In each case, we chose a basic set of parameters to achieve bistability in the absence of oscillatory feedback.  Feedback was then introduced and parameters were varied to determine whether a given model variant is consistent with experimental manipulations described by \cite{JPark2016}.

\section{Model equations}

\subsection{Model 1, ECM signaling equation analysis}
Consider the ECM signaling equations Eqs.~\eqref{gen-nd} with $A_{E}(R_{1,2}), L_E(\rho_{1,2})$ taken to be parameters. 
The equations for the ECM are then
\[
\frac{dE_{k}}{dt}=\epsilon \left[(B_E+ A_E E_{k})  -   E_k\left(L_E E_{k} + l_c E_{j}  \right)   \right],  \quad  j \neq k. \label{ECMa}
\]
This is the Lotka-Volterra species-competition model, briefly analyzed here. This system admits four possible solutions $s_{1}=(0,0)$, $s_2=(0,E_2^{*})$, $s_3=(E_1^*,0)$, and $s_4=(E_1^{**},E_2^{**})$ where
\begin{equation}
E^{*}_{1,2}=\frac{A_E(R_{1,2})}{L_E(\rho_{1,2})}, \qquad \textnormal{and} \qquad E^{**}_{1,2}= \frac{A_E(R_{1,2}) L_E(\rho_{2,1})-A_E(\rho_{2,1}) l_c}{L_E(\rho_{1,2}) L_E(\rho_{2,1})-l_c^2}.
\end{equation}
Then $s_{2,3}$ are associated with two polarized states while $s_4$ is an apolar co-existence state. In region I, the only stable solution is $s_4$. In regions II and III, solutions $s_2$ and $s_3$ respectively are stable. In Regime IV, only $s_1$ is stable.

\subsection{Model 1a}

Using $R_I=\rho_I=1$ as before, and taking $n=1$, $B_E=0$ and a linear assumption for $b_\rho$ as in Table \ref{table:equations}, we arrive at the model equations
\begin{subequations}\label{Model2aAppdx}
 \begin{align} 
\frac{dR_k}{dt} &=  \frac{b_R}{1+\rho_k} R_I - \delta R_k, \\ 
\frac{d \rho_k}{dt} &=  \frac{(k_{E} + \bar{\gamma}_E E_k)}{1+ R_k} \rho_I - \rho_k , \\
\frac{dE_{k}}{dt}&=\epsilon \left[ (k_R + \bar{\gamma}_R R_k)E_{k}   -   ((k_{\rho} + \bar{\gamma}_{\rho} \rho_k) E_{k}^2 + l_c E_{j} E_{k} ) \right], 
\end{align}
for $k=1,2$ representing quantities in the two lamellipods and $j \neq k$ representing quantities in the other.
\end{subequations}

\subsection{Model 1b}

The equations for Model 1b are the same as those of Model 1a, but with conservation of total GTPases, $R_I=R_T-R_1-R_2$ and $\rho_I=\rho_T-\rho_1-\rho_2$ as in 
\eqref{GTPaseConserv}.

Technically, the transition from polarized to oscillatory behavior shown in Figure~\ref{fig:2} results from a bifurcation to heteroclinic cycles rather than a canonical Hopf bifurcation.  (But the conclusions are similar.)
In the oscillatory regime of this model, there are two unstable polarized states.  The oscillation is a heteroclinic cycle connecting these two states.  As the bifurcation is approached from the oscillatory side, the period of  oscillation becomes infinitely long, and at the bifurcation the system spends infinite time at one or the other polarized states, representing the transition to stability.  

\subsubsection{Model 1b parameter set:}
$k_R=1, l_c=1, k_{\rho}=0.1, k_{E}=0, \bar{\gamma}_E=5,  \epsilon=10, b_R=0.5, \delta=1,  \bar{\gamma}_R=0.75, \bar{\gamma}_{\rho}=1,n=1, R_T=\rho_T=1$.

\subsection{Model 1c}
This model augments Model 1b to consider the influence of Rac and Rho effector molecules in the protrusion and contraction process. Here $w_k$ represents effectors of Rac and $c_k$ effectors of Rho in lamellipod $k$. We assume that these directly influence protrusion and contraction terms ($B_E$ and $L_E$) and that their dynamics are described by linear ODE's. We retain the same Rac and Rho equations as in Model 1b and include the following to describe dynamics of $E_k, w_k, c_k$.
\begin{subequations} \label{full-model-2c}
\begin{align}
\frac{d E_k}{d t} & =  \epsilon \left[(k_R+w_k)E_k  -  ((k_{\rho} + c_k) E_k^2 +\beta E_jE_k) \right], \quad j \ne k,\\
\frac{d w_k}{d t} &=\epsilon_2 ( \gamma_R R_k - w_k) ,\\
\frac{d c_k}{d t} &=\epsilon_2 ( \gamma_{\rho} \rho_k - c_k).
\end{align}
\end{subequations}
Here, $1/\epsilon_2$ is a timescale variable associated with the speed of these reactions. Small $\epsilon_2$ indicates $w_k$ and $c_k$ are slow variables. 

\begin{figure}[htbp] 
  \centerline{ 
\includegraphics[scale=.75]{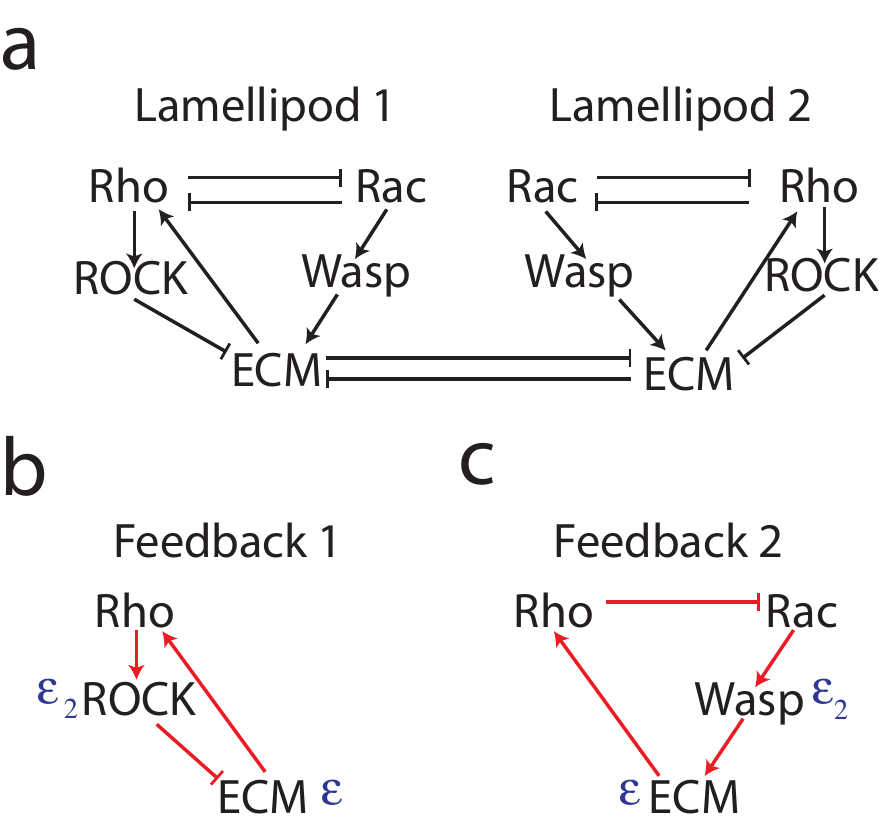}}
\caption{
Schematic diagram for Model {\bf 1}c. Similar to Figure \ref{fig:1}a except with the explicit inclusion of intermediate effectors such as Rock and Wasp / Wave.  \textbf{(a)} Full model describing the proposed interactions between the ECM and GTPase signalling.  \textbf{(b,c)} Diagram of the two negative feedback loops embedded in this model that are capable of producing oscillations.   $\epsilon_1, \epsilon_2, \epsilon_3$ are rate constants for GTPase dynamics, Rock / Wasp dynamics, and ECM dynamics respectively, see \eqref{full-model-2c}.  }
\label{fig:5}
\end{figure} 

This reformulation leads to three explicit timescales representing GTPase dynamics $O(1)$, dynamics of GTPase effectors $O(\epsilon_2)$, and dynamics of the ECM $O(\epsilon)$.  Model {\bf 1}c (Eqs.~\eqref{full-model-2c}) asymptotically reduces to Model {\bf 1}b provided the GTPase effector dynamics (e.g. ROCK / WASP) are fast ($\epsilon_2  \gg \epsilon$). In this case a quasi steady state approximation reduces Model {\bf 1}c to Model {\bf 1}b and bifurcation analysis of the full Model {\bf 1}c equations (not shown) closely matches that of Figure \ref{fig:2}.  In that case, Model 1c has the same problem that GTPase dynamics are required to be slow to generate oscillations. When $\epsilon_2 \ll \epsilon$, the effectors determine the timescale of feedbacks, and consequently, GTPases need not be slower than ECM dynamics (e.g. it can be the case that $\epsilon < 1$).     

\section{Model 2}

We set $R_I=\rho_I=1$, and Hill coefficients $n=3$, and use the assumed $b_{\rho}(E_k)$ expression. Setting $A_E=0$, and using assumed forms of $B_E$ and $L_E$, we get (for $k=1,2$ and $j \ne k$):

\begin{subequations}\label{Model1Appdx}
 \begin{align} 
\frac{dR_k}{dt} &=  \frac{b_R}{1+\rho_k^3} R_I - \delta R_k, \\
\frac{d \rho_k}{dt} &=  \left( k_E+\gamma_E \frac{E_k^3}{E_0^3+E_k^3} \right)\frac{1}{1+ R_k^3} \rho_I - \rho_k, \\
\frac{dE_{k}}{dt}&=\epsilon \left[\left(k_R+ \gamma_R \frac{R_k^3}{R_0^3+R_k^3}\right)- E_{k}^2\left(k_{\rho}+\gamma_{\rho} \frac{\rho_k^3}{\rho_0^3+\rho_k^3}\right) - l_c E_{j} E_{k}    \right].  
\end{align}
\end{subequations}

\subsubsection{Model 2 parameter set}
$b_R=2, n=3, \delta=1, k_E=k_R=k_{\rho}=0, \gamma_E=5, R_0=0.85, \rho_0=0.85, E_0=1, l_c=1, \epsilon=0.1, R_I=\rho_I=1$.

\section{Model 3}

We use the conservation of GTPases \eqref{GTPaseConserv} combined with the bistable GTPase model (2)  to obtain the equations:


\begin{subequations}\label{ModeldAppdx}
 \begin{align} 
\frac{dR_k}{dt} &=  \frac{b_R}{1+\rho_k^3} (R_T-R_1-R_2) - \delta R_k, \\
\frac{d \rho_k}{dt} &=  \left( k_E+\gamma_E \frac{E_k^3}{E_0^3+E_k^3} \right)\frac{1}{1+ R_k^3} (\rho_T-\rho_1-\rho_2) - \rho_k .
\end{align}
\begin{equation}
\frac{dE_{k}}{dt}=\epsilon \left[\left(k_R+ \gamma_R \frac{R_k^3}{R_0^3+R_k^3}\right)- E_{k}^2\left(k_{\rho}+\gamma_{\rho} \frac{\rho_k^3}{\rho_0^3+\rho_k^3}\right) - l_c E_{j} E_{k}    \right]. 
\end{equation}
\end{subequations}

\subsubsection{Model 3 parameter set}
The base parameters used for this model are $R_T= \rho_T=2, \delta=1,   b_R=5, R_0=  \rho_0=0.85, E_0=1, n=3, k_E=3, \gamma_E=2,  \epsilon=0.1, k_{\rho }=0.2, k_{R}=0.2, \gamma_R=0.3, l_c=0$.

\section{Summary of terms and types of models}


\begin{table}
\begin{tabular}{|p{2.5cm} | p{5.5cm} | p{5cm} |}\hline
     & Source of Polarity & Source of anti-phase oscillations \\ \hline
   Lamellipod Competition (1)   &     Bistable lamellipod coupling                & Fast ECM / Slow GTPase feedback \\
      							 &	    Conserved GTPase with MM kinetics & Bistable lamellipod coupling \\ \hline
    Bistable GTPase (2)            &     Bistable GTPase model                & Slow ECM / Fast GTPase feedback \\
      							&	    Monostable lamellipod coupling   & Monostable lamellipod coupling \\ \hline
                                                 &     Conservative GTPase model        & Slow ECM / Fast GTPase feedback \\
    Hybrid  (3)                           &     with Hill kinetics                             &  \\
      							&	    Monostable lamellipod coupling   & Monostable lamellipod coupling \\ \hline

\hline
\end{tabular}
\caption{\label{Tab:Source} We describe the qualitative elements of each model that give rise to polarity and antiphase lamellipod oscillations, respectively. MM: Michaelis Menten kinetics. Numbers indicate the model designation used in the text.
}
\end{table}

\begin{table}
\begin{tabular}{| p{3cm} | c | c |}\hline
     & GTPase Model & Lamellipod coupling \\ \hline
   Lamellipod Competition (1)   &     Michaelis Menten +               & Bistable, Species competition \\
      							 &	   conservation                         & type equations \\ \hline
    Bistable GTPase (2)            &     Bistable Hill kinetics               & Monostable \\ \hline
    Hybrid  (3)                           &     Hill kinetics + conservation    &  Monostable \\  \hline

\hline
\end{tabular}
\caption{\label{Tab:Elements} Terms that represent the GTPase and lamellipod coupling in each sub-model. 
}
\end{table}

\begin{table}
\begin{tabular}{|p{2.75cm} | c | c |  c |c |}\hline
Parameter definition& Notation&Model 1&Model 2&Model 3 \\  \hline
Rac activation rate &$A_{R}(\rho)$&$\displaystyle \frac{\hat{b}_R}{({\hat \rho_0}+\rho)}$&$\displaystyle \frac{\hat{b}_R}{({\hat \rho_0}^3+\rho^3)}$&$\displaystyle \frac{\hat{b}_R}{({\hat \rho_0}^3+\rho^3)}$\\
Rho activation rate &$A_{\rho}(R)$&$\displaystyle \frac{\hat{b}_\rho}{({\hat R_0}+ R)}$ &$\displaystyle \frac{\hat{b}_\rho}{({\hat R_0}^3+ R^3)}$& $\displaystyle \frac{\hat{b}_\rho}{({\hat R_0}^3+ R^3)}$\\
ECM effect on Rho activation &$b_{\rho} (E_k)$& $k_{E} + \bar{\gamma}_E E_k$ & $\displaystyle k_E+\gamma_E \frac{E_k^3}{E_0^3+E_k^3}$ &$\displaystyle k_E+\gamma_E \frac{E_k^3}{E_0^3+E_k^3}$ \\
&&&&\\
\hline
&&&&\\
Protrusion; ECM growth rate&$P(R_k,E_{k})$&&& \\
- - Basal Rac-dependent term &$B_E(R_k)$& 0 & $\displaystyle k_R+ \gamma_R \frac{R_k^3}{R_0^3+R_k^3}$ & $\displaystyle k_R+ \gamma_R \frac{R_k^3}{R_0^3+R_k^3}$ \\
- - ECM autoamplif. term & $A_E(R_k)$& $k_R + \bar{\gamma}_R R_k$ & 0 & 0 \\
&&&&\\
\hline
&&&&\\
Rho dependent contraction & $L_E(\rho_k)$ & $k_{\rho} + \bar{\gamma}_{\rho} \rho_k$ &  $\displaystyle k_{\rho}+\gamma_{\rho} \frac{\rho_k^3}{\rho_0^3+\rho_k^3}$ & $\displaystyle k_{\rho}+\gamma_{\rho} \frac{\rho_k^3}{\rho_0^3+\rho_k^3}$\\
&&&&\\
\hline
&&&&\\
Total GTPase conserved& $G_T$ & (2a) no; (2b)\checkmark & no &\checkmark \\
&&&&\\
\hline
\end{tabular}
\caption{\label{Tab:ModelTerms2}Terms and notation used in model equations.
Model 1: GTPase bistability, Model 2: Lamellipod competition, Model 3: Hybrid model. Parameters $\bar{\gamma}_{E,R,\rho}$ have similar meanings as $\gamma_{E,R,\rho}$ in Models 1,3 but carry distinct units to accommodate linear versus Hill function kinetic terms.
}
\label{table:equations}
\end{table}

\end{document}